\documentclass[]{article}
\usepackage[]{graphicx}
\begin{document}

\title{Propagation of localized optical waves in media with dispersion, in dispersionless media and in vacuum.
Low diffractive regime}

\author{Lubomir M. Kovachev\\
Institute of Electronics, Bulgarian Academy of Sciences,\\
Tzarigradcko shossee 72,1784 Sofia, Bulgaria}
\maketitle
\date{}

\begin{abstract}
We present a systematic study on linear propagation of ultrashort
laser pulses in media with dispersion, dispersionless media and
vacuum. The applied method of amplitude envelopes gives the
opportunity to estimate the limits of slowly warring amplitude
approximation and to describe an amplitude integro-differential
equation, governing the propagation of optical pulses in single
cycle regime. The well known slowly varying amplitude equation and
the amplitude equation for vacuum are written in dimensionless form.
Three parameters are obtained defining different linear regimes of
the optical pulses evolution. In contrast to previous studies we
demonstrate that in femtosecond region the nonparaxial terms are not
small and can dominate over transverse Laplacian. The normalized
amplitude nonparaxial equations are solved using the method of
Fourier transforms. Fundamental solutions  with  spectral kernels
different from Fresnel one are found. One unexpected new result is
the relative stability of light pulses with spherical and spheroidal
spatial form, when we compare their transverse enlargement with the
paraxial diffraction of lights beam in air. It is important to
emphasize here the case of light disks, i.e. pulses whose
longitudinal size is small with respect to the transverse one, which
in some partial cases are practically diffractionless over distances
of thousand kilometers. A new formula which calculates the
diffraction length of optical pulses is suggested.
\end{abstract}

\section{Introduction}

For long time few picosecond or femtosecond (fs) optical pulses with
approximately equal duration in the $x$, $y$ and $z$ directions
(Light Bullets or LB), and fs optical pulses with relatively large
transverse and small longitudinal size (Light Disks or LD) are used
in the experiments. The evolution of so generated LB and LD in
linear or nonlinear regime is quite different from the propagation
of  light beams and they have drawn the researchers' attention with
their unexpected dynamical behavior. For example, self-channeling of
femtosecond pulses with power little above the critical for
self-focusing \cite{BRAUN} and also below the nonlinear collapse
threshold \cite{RUIZ} (linear regime) in air, was observed. This is
in contradiction with the well known self-focusing and diffraction
of an optical beam in the frame of paraxial optics. Various
unidirectional propagation equations have been suggested to be found
stable pulse propagation mainly in nonlinear regime (see e.g.
Moloney and Kolesik \cite{MOLKOL}, Couairon and Mysyrowicz
\cite{COUAI}, Chin at all. \cite{CHIN}, for a review). The basic
studies in this field started with the so called spatio-temporal
nonlinear Schr\"{o}dinger equation (NSE) which is one compilation
between paraxial approximation, the group velocity dispersion (GVD)
and nonlinearity \cite{SIL, PET, KOPR, KIV}. The influence of
additional physical effects were  studied by adding different terms
to this scalar model as small nonparaxiality \cite{AKH, FIB}, plasma
defocussing, multiphoton ionization and vectorial generalizations.
It is not hard to see that for pulses with low intensity (linear
regime) in air and gases the additional terms as GVD and others
become small and the basic model can be reduced to paraxial
equation. This is the reason diffraction of a low intensity optical
pulse governed by this model on several diffraction length to be
equal to diffraction of a laser beam. On other hand, the
experimentalists have discussed for a long time that in their
measurements the diffraction length of an optical pulse is not equal
to this of a laser beam $z^{beam}_{diff}=k_0r^2_{\perp}$, even when
additional phase effects of lens and other optical devices can be
reduced. Here $k_0$ denotes laser wave-number and $r^2_{\perp}$
denotes the beam waist. Thus exist one deep difference between the
existing models in linear regime, predicting paraxial behavior in
gases, and the real experiments.

The purpose of this work is to perform  a systematic study of linear
propagation of ultrashort optical pulses in media with dispersion,
dispersionless media and vacuum and to suggest a model which is more
close to the experimental results. In addition, there are several
particular problems under consideration in this paper.

The first one is to obtain (not slowly varying) amplitude envelope
equation in media with  dispersion  governing the evolution of
optical pulses in single-cycle regime. This problem is natural in
femtosecond region where the optical period of a pulse is of order
$2-3$ fsec. The earliest model for pulses in single-cycle regime
suggested by Brabec and Krausz \cite{BRAB} is obtained after Taylor
expansion of wave vector $k^2(\omega)$ about $\omega_0$. It is  easy
to show \cite{KAMEN} that this expansion diverge in solids for
single-cycle pulses. The higher order dispersion terms start to
dominate and the series can not cut off. This is the reason more
carefully and accurately to derive the envelope equation before
using Taylor series.  In this way we obtain an integro-differential
envelope equation where no Taylor expansion of the wave vector
$k^2(\omega)$, governing evolution of single cycle pulses in solids.

The second problem is to investigate more precisely the slowly
varying envelope equations governing the evolution of optical pulses
with high number of harmonics under the envelope. The slowly varying
scalar Nonlinear Envelope Equation (NEE) is derived in many books
and papers \cite{KARP, JAIN, MN,  CHRIS, AKH, BOYD, KOV}. After the
deriving of the NEE, most of the authors use a standard procedure to
neglect the nonparaxial terms as small ones. Only some partial
nonparaxial approximations in free space \cite{CHRISTOV, AKH, FIB}
and optical fibers \cite{MENYUK} were studied. In \cite{KOV1} we
rewrite the NEE in dimensionless form and estimate the influence of
the different linear and nonlinear terms on the evolution of optical
pulses. We found that both nonparaxial terms  in NEE, second
derivative in propagation direction and second derivative in time
with $1/v^2$ coefficient, are not small corrections. In fs region
they are of same order as transverse Laplacian or start to dominate.
These equations with (not small) nonparaxial terms  are solved in
linear regime \cite{KOV1} and investigated numerically in nonlinear
\cite{KOV2}. In this paper we include  GVD term in the nonparaxial
model and study also the envelope equation of electrical field in
vacuum and dispersionless media. It is important to note that the
Vacuum Linear Amplitude Equation (VLAE) is obtained without any
expansion of the wave vector. That is why it work also for pulses in
single-cycle regime (subfemto and attosecond pulses).

Last but not least the nonparaxial equations for media with
dispersion, dispersionless media and vacuum are solved in linear
regime and new fundamental solutions, including the GVD, are found.
The solutions of these equations predict new diffraction length for
optical pulses $z^{pulse}_{diff}=k_0^2r^4_{\perp}/z_0$, where $z_0$
is the longitudinal spatial size of the pulse (the spatial analog of
the time duration $t_0$; $z_0=vt_0$; v is group velocity). In case
of fs propagation in gases and vacuum we demonstrate by these
analytical and numerical solutions a significant decreasing of the
diffraction enlargement in respect to paraxial beam model and a
possibility to reach practically diffraction-free regime.

\section{From Maxwell's equations of a source-free, dispersive,\\
nonlinear Kerr type medium to the amplitude equation}

The propagation of ultra-short laser pulses in isotropic media, can
be characterized by the following dependence of the polarization of
first $\vec P_{lin}$ and third $\vec P_{nl}$ order on the electrical
field $\vec E$:

\begin{eqnarray}
\label{eq6} \vec P_{lin} = \int\limits_{-\infty}^{t} {\left(\delta
(\tau-t) +4\pi\chi^{\left(1\right)} \left(\tau-t\right)\right)\vec
E\left(\tau, r\right)}d\tau =\nonumber\\
\int\limits_{-\infty}^{t} {\varepsilon \left(\tau-t\right)\vec
E\left(\tau, x, y, z\right)}d\tau,
\end{eqnarray}

\begin{eqnarray}
\label{pnl} \vec P_{nl}^{(3)} =3\pi
\int\limits_{-\infty}^{t}\int\limits_{-\infty}^{t}\int\limits_{-\infty}^{t}
\chi^{\left(3\right)}
\left(\tau_1-t,\tau_2-t,\tau_3-t\right)\nonumber\\
\times\left(\vec{E}(\tau_1,r)\cdot\vec E^{*}(\tau_2,r)\right)\vec
E(\tau_3,r) d\tau_1d\tau_2d\tau_3,
\end{eqnarray}
where $\chi^{(1)}$ and $\varepsilon$ are the linear electric
susceptibility and the dielectric constant, $\chi^{(3)}$ is the
nonlinear susceptibility of third order, and we denote $r=(x,y,z)$.
We use the expression of the nonlinear polarization (\ref{pnl}), as
we will investigate only linearly or only circularly polarized light
and in addition we neglect the third harmonics term. The Maxwell's
equations in this case becomes:

\begin{eqnarray}
\label{eq1}
\nabla \times \vec E =
- \frac{1}{c}\frac{\partial \vec B}{\partial t},
\end{eqnarray}

\begin{eqnarray}
\label{eq2}
\nabla \times \vec H =
  \frac{1}{c}\frac{\partial \vec D}{\partial t},
\end{eqnarray}

\begin{equation}
\label{eq3}
\nabla \cdot \vec D = 0,
\end{equation}

\begin{equation}
\label{eq4}
\nabla \cdot \vec B = \nabla \cdot \vec H = 0,
\end{equation}

\begin{equation}
\label{eq5} \vec B = \vec H,\ \vec D = \vec P_{lin}  +  \vec P_{nl},
\end{equation}
where $\vec E$ and $\vec H$ are the electric and magnetic fields
strengths, $\vec D$ and $\vec B$ are the electric and magnetic
inductions. We should point out here that these equations are valid
when the time duration of the optical pulses $t_0$ is greater than
the characteristic response time of the media $\tau_0$
($t_0>>\tau_0$), and also when the time duration of the pulses is of
the order of time response of the media ($t_0\leq\tau_0$). Taking
the curl of equation (\ref{eq1}) and using (\ref{eq2}) and
(\ref{eq5}), we obtain:

\begin{eqnarray}
\label {eq8} \nabla \left(\nabla\cdot\vec E\right) -\Delta\vec E=
-\frac{1}{c^2}\frac{\partial^2\vec D}{\partial t^2},
\end{eqnarray}
where $\Delta \equiv \nabla ^2$ is the Laplace operator. Equation
(\ref{eq8}) is derived without using the third Maxwell's equation.
Using equation (\ref{eq3}) and the expression for the linear and
nonlinear polarizations (\ref{eq6}) and (\ref{pnl}), we can estimate
the second term in equation (\ref{eq8}) for arbitrary localized
vector function of the electrical field. It is not difficult  to
show that for localized functions in nonlinear media with and
without dispersion  $\nabla \cdot\vec {E}\cong 0$ and we can write
equation (\ref{eq8}) as follows:
\begin{eqnarray}
\label {lap} \Delta\vec E= \frac{1}{c^2}\frac{\partial^2\vec
D}{\partial t^2}.
\end{eqnarray}
We will now replace the electrical field in linear and nonlinear
polarization on the right-hand side of (\ref{lap}) with it's Fourier
integral:

\begin{eqnarray}
\label {F1} \vec E\left(r,t\right)= \int\limits_{-\infty}^{+\infty}
\hat{\vec E}\left(r,\omega\right) \exp{\left(-i\omega
t\right)}d\omega,
\end{eqnarray}
where with $\hat{\vec E}\left(r,\omega\right)$ we denote the time
Fourier transform of the electrical field. We thus obtain:

\begin{eqnarray}
\label{lapint} \Delta\vec E= \frac{1}{c^2}\frac{\partial^2}{\partial
t^2} \left(\int\limits_{-\infty}^{t}
\int\limits_{-\infty}^{\infty}\varepsilon(\tau-t) \hat{\vec
E}\left(r,\omega\right)\exp{(-i\omega\tau)}d\omega d\tau\right)
+\nonumber\\
\\
\frac{3\pi}{c^2}\frac{\partial^2}{\partial t^2}
\int\limits_{-\infty}^{t}\int\limits_{-\infty}^{t}
\int\limits_{-\infty}^{t}\int\limits_{-\infty}^{\infty}\chi^{\left(3\right)}
\left(\tau_1-t,\tau_2-t,\tau_3-t\right)\left|\hat{\vec
E}\left(r,\omega\right)\right|^2 \hat{\vec
E}\left(r,\omega\right)\nonumber\\
\times\exp{\left(-i\left(\omega(\tau_1-\tau_2+\tau_3)\right)\right)}
d\omega d\tau_1d\tau_2d\tau_3.\nonumber
\end{eqnarray}
The causality principle imposes the following conditions on the
response functions:

\begin{eqnarray}
\label{caus} \varepsilon(\tau-t)=0;\ \chi^{\left(3\right)}
\left(\tau_1-t,\tau_2-t,\tau_3-t\right)=0, \nonumber\\
 \tau-t>0;\ \tau_i-t>0;\ i=1,2,3.
\end{eqnarray}
That is why we can extend the upper integral boundary to infinity
and use the standard Fourier transform \cite{MN}:

\begin{eqnarray}
\label{intchi1} \int\limits_{-\infty}^{t}
{\varepsilon(\tau-t)\exp{(-i\omega\tau)}d\tau}=\int\limits_{-\infty}^{+\infty}
{\varepsilon(\tau-t)\exp{(-i\omega\tau)}d\tau},
\end{eqnarray}

\begin{eqnarray}
\label{intchi3}
\int\limits_{-\infty}^{t}\int\limits_{-\infty}^{t}\int\limits_{-\infty}^{t}
\chi^{\left(3\right)} \left(\tau_1-t,\tau_2-t,\tau_3-t\right)
d\tau_1d\tau_2d\tau_3=\nonumber\\
\int\limits_{-\infty}^{+\infty}\int\limits_{-\infty}^{+\infty}\int\limits_{-\infty}^{+\infty}
\chi^{\left(3\right)} \left(\tau_1-t,\tau_2-t,\tau_3-t\right)
d\tau_1d\tau_2d\tau_3.
\end{eqnarray}
The spectral representation of the linear optical susceptibility
$\hat{\varepsilon_{0}}(\omega)$ is connected to the non-stationary
optical response function by the following Fourier transform:

\begin{eqnarray}
\label{chi1} \hat{\varepsilon}(\omega)\exp{(-i\omega t)}
=\int\limits_{-\infty}^{+\infty}
{\varepsilon(\tau-t)\exp{(-i\omega\tau)}d\tau}.
\end{eqnarray}
The expression for the spectral representation of the non-stationary
nonlinear optical susceptibility $\hat{\chi}^{(3)}$ is similar :

\begin{eqnarray}
\label{chi3} \hat{\chi}^{(3)}(\omega)\exp{(-i\omega t)}=
\int\limits_{-\infty}^{+\infty}\int\limits_{-\infty}^{+\infty}
\int\limits_{-\infty}^{+\infty}\chi^{\left(3\right)}
\left(\tau_1-t,\tau_2-t,\tau_3-t\right)\nonumber\\
\times\exp{\left(-i\left(\omega(\tau_1-\tau_2+\tau_3)\right)\right)}
d\tau_1d\tau_2d\tau_3.
\end{eqnarray}
Thus, after brief calculations, equation (\ref{lapint}) can be
represented as

\begin{eqnarray}
\label{lapomeg} \Delta\vec E=-\int\limits_{-\infty}^{\infty}
\frac{\omega^2\hat{\varepsilon}(\omega)}{c^2} \hat{\vec
E}\left(r,\omega\right)\exp{(-i\omega t)}d\omega
\nonumber\\
+
\int\limits_{-\infty}^{\infty}\frac{\omega^2\hat{\chi}^{\left(3\right)}
\left(\omega\right)}{c^2}\left|\hat{\vec
E}\left(r,\omega\right)\right|^2 \hat{\vec
E}\left(r,\omega\right)\exp{\left(-i\left(\omega t\right)\right)}
d\omega.
\end{eqnarray}
We now define the square of the linear $k^2$ and the generalized
nonlinear $\hat{k}_{nl}^2$ wave vectors, as well as the nonlinear
refractive index $n_2$ with the expressions:

\begin{eqnarray}
k^2=\frac{\omega^2\hat{\varepsilon}\left(\omega\right)}{c^2},
\end{eqnarray}

\begin{eqnarray}
\label{gnlv}
\hat{k}_{nl}^2=\frac{3\pi\omega^2\hat{\chi}^{(3)}\left(\omega\right)}{c^2}=k^2n_2,
\end{eqnarray}
where
\begin{eqnarray}
n_2(\omega)=\frac{3\pi\hat{\chi}^{(3)}\left(\omega\right)}{\hat{\varepsilon}\left(\omega\right)}.
\end{eqnarray}
The connection  between the usual dimensionless nonlinear wave
vector $k_{nl}^2$ and the generalized one (\ref{gnlv}) is:
$k_{nl}^2=\hat{k}_{nl}^2\left|\hat{\vec
E}\left(r,\omega\right)\right|^2$. In terms of these quantities,
equation (\ref{lapomeg}) can be expressed by:

\begin{eqnarray}
\label{lapk} \Delta\vec E=-\int\limits_{-\infty}^{\infty}
k^2(\omega) \hat{\vec E}\left(r,\omega\right)\exp{(-i\omega
t)}d\omega
\nonumber\\
-
\int\limits_{-\infty}^{\infty}k^2(\omega)n_2(\omega)\left|\hat{\vec
E}\left(r,\omega\right)\right|^2 \hat{\vec
E}\left(r,\omega\right)\exp{\left(-i\left(\omega t\right)\right)}
d\omega.
\end{eqnarray}
Let us introduce here the amplitude function $\vec A(r,t)$ for the
electrical field $\vec E(r,t)$:

\begin{eqnarray}
\label{a1}
 \vec E\left(x,y,z,t\right)=\vec
{A}\left(x,y,z,t\right)\exp{\left(i(k_0z-\omega_0t)\right)},
\end{eqnarray}
where $\omega_0$ and $k_0 $ are the carrier frequency and the
carrier wave number of the wave packet. The writing of the amplitude
function in this form means that we consider  propagation only in
$+z$ -direction and neglect the opposite one. Let us write here also
the Fourier transform of the amplitude function $\hat{\vec
A}(r,\omega-\omega_0)$:

\begin{eqnarray}
\label {FA} \vec A\left(r,t\right)= \int\limits_{-\infty}^{+\infty}
\hat{\vec A}\left(r,\omega-\omega_0\right)
\exp{\left(-i(\omega-\omega_0) t\right)}d\omega,
\end{eqnarray}
and the following relation between the Fourier transform of the
electrical field and the Fourier transform of the amplitude
function:

\begin{eqnarray}
\label{hata}
 \hat{\vec E}\left(r,\omega\right)\exp(-i\omega t)=\nonumber\\
 \exp\left(-i\left(k_0z-\omega_0 t\right)\right)
 \hat{\vec
 {A}}\left(r,\omega-\omega_0\right)\exp{\left(i(\omega-\omega_0)t\right)},
\end{eqnarray}
Since we investigate optical pulses, we assume that the amplitude
function and its Fourier expression are time - and
frequency-localized. Substituting (\ref{a1}),(\ref{FA}) and
(\ref{hata}) into equation (\ref{lapk}) we finally obtain the
following nonlinear integro-differential amplitude equation:

\begin{eqnarray}
\label{ampk} \Delta\vec A(r,t) + 2ik_0\frac{\partial\vec
A(r,t)}{\partial
z}- k_0^2\vec A(r,t)=\nonumber \\
\\
-\int\limits_{-\infty}^{\infty}
k^2(\omega)\left(1+n_2(\omega)\left|\hat{\vec
A}\left(r,\omega-\omega_0\right)\right|^2 \right) \hat{\vec
A}\left(r,\omega-\omega_0\right)\exp{(-i(\omega-\omega_0)
t)}d\omega\nonumber
\end{eqnarray}
Equation (\ref{ampk}) was derived with only one restriction, namely,
that the amplitude function and its Fourier expression are localized
functions. That is why, if we know the analytical expression of
$k^2(\omega)$ and $n_2(\omega)$, the Fourier integral on the right-
hand side of (\ref{ampk}) is a finite integral away from resonances.
In this way we can also investigate optical pulses with time
duration $t_0$ of the order of the optical period
$T_0=2\pi/\omega_0$. Generally, using the nonlinear
integro-differential amplitude equation (\ref{ampk}) we can also
investigate wave packets with time duration of the order of the
optical period, as well as wave packets with a large number of
harmonics under the pulse. The nonlinear integro-differential
amplitude equation (\ref{ampk}) can be written as a nonlinear
differential equation for the Fourier transform of the amplitude
function $\hat{\vec A}$, after we apply the time Fourier
transformation (\ref{FA}) to the left-hand side of (\ref{ampk}) :

\begin{eqnarray}
\label{ampf} \Delta\hat{\vec A}\left(r,\omega-\omega_0\right)+
2ik_0\frac{\partial\hat{\vec
A}\left(r,\omega-\omega_0\right)}{\partial z}\nonumber \\
+\left(\left(1+n_2(\omega)\left|\hat{\vec
A}\left(r,\omega-\omega_0\right)\right|^2
\right)k^2(\omega)-k_0^2(\omega_0)\right) \hat{\vec
A}\left(r,\omega-\omega_0\right)=0.
\end{eqnarray}
We should note here the well-known fact that the Fourier component
of the amplitude function in equation (\ref{ampf}) depends on the
spectral difference $\triangle\omega=\omega-\omega_0$, rather than
on the frequency, as is the case for the electrical field.

\section{From amplitude equation to the slowly varying envelope approximation (SVEA)}

Equation (\ref{ampk}) is obtained without imposing any restrictions
on the square of the linear $k^2(\omega)$ and generalized nonlinear
$\hat{k}^2_{nl}=k^2(\omega)n_2(\omega)$ wave vectors. To obtain
SVEA, we will restrict our investigation to the cases when it is
possible to approximate $k^2$ and $\hat{k}_{nl}^2$ as a power series
with respect to the frequency difference $\omega-\omega_0$ as:

\begin{eqnarray}
\label{a9}
 k^2\left(\omega\right)=
\frac{\omega ^2\hat{\varepsilon_0}\left(\omega\right)}{c^2} =
 k^{2}\left(\omega_0\right) +
\frac{\partial\left(k^2\left(\omega_0\right)\right)}{\partial\omega_0}
\left(\omega-\omega_0\right) \nonumber\\
+ \frac{1}{2}
\frac{\partial^2\left(k^2\left(\omega_0\right)\right)}{\partial
\omega_0^2} \left(\omega-\omega_0\right)^2 + ...,
\end{eqnarray}
\begin{eqnarray}
\label{nla9}
 \hat{k}_{nl}^2\left(\omega\right)=
\frac{\omega ^2\hat{\chi}^{(3)}\left(\omega\right)}{c^2} =
\hat{k}_{nl}^{2} \left(\omega_0\right) +
\frac{\partial\left(\hat{k}_{nl}^2\left(\omega_0\right)\right)}{\partial\omega_0}
\left(\omega-\omega_0\right)+...
\end{eqnarray}
To obtain SVEA in second approximation to the linear dispersion and
in first approximation to the nonlinear dispersion, we must cut off
these series to the second derivative term for the linear wave
vector and to the first derivative term for the nonlinear wave
vector. This is possible only if the series (\ref{a9}) and
(\ref{nla9}) are strongly convergent. Then, the main value in the
Fourier integrals in equation (\ref{ampk}) yields the first and
second derivative terms in (\ref{a9}), and the zero and first
derivative terms in (\ref{nla9}). The first term in (\ref{a9})
cancels the last term on the left-hand side of equation
(\ref{ampk}). The convergence of the series (\ref{a9}) and
(\ref{nla9}) for spectrally limited pulses propagating in the
transparent UV and optical regions of solids materials, liquids and
gases, depends mainly on the number of harmonics under the pulses
\cite{KAMEN}. For wave packets with more than 10 harmonics under the
envelope, the series (\ref{a9}) is strongly convergent, and the
third derivative term (third order of dispersion) is smaller than
the second derivative term (second order of dispersion) by three to
four orders of magnitude for all materials. In this case we can cut
the series to the second derivative term in (\ref{a9}), as the next
terms in the series contribute very little to the Fourier integral
in equation (\ref{ampk}). When there are $2-6$ harmonics under the
pulse, the series (\ref{a9}) is weakly convergent for solids and
continue to be strongly convergent for gases. Then for solids we
must take into account the dispersion terms of higher orders as
small parameters. In the case of wave packets with only one or two
harmonics under the envelope, propagating in solids, the series
(\ref{a9}) is divergent. This is the reason why the SVEA does not
govern the dynamics of wave packets with time duration of the order
of the optical period in solids. Substituting the series (\ref{a9})
and (\ref{nla9}) in (\ref{ampk}) and bearing in mind the expressions
for the time-derivative of the amplitude function, the SVEA of
second order with respect to the linear dispersion and first order
with respect to the nonlinear dispersion is expressed in the
following form:

\begin{eqnarray}
\label{sve} \Delta\vec A + 2ik_0\frac{\partial\vec A}{\partial z}+
2ik_0k'\frac{\partial\vec A}{\partial
t}=\nonumber\\
\left(k_0k"+k'^2\right)\frac{\partial^2\vec A}{\partial
t^2}-\hat{k}_{0nl}^2\left|\vec A\right|^2 \vec
A-2i\hat{k}_{0nl}\hat{k}'_{0nl}\frac{\partial\left|\vec A\right|^2
\vec A}{\partial t},
\end{eqnarray}
where $k_0= k(\omega_0)$ and $\hat{k}_{0nl}^2=\hat{k}_{nl}^{2}
\left(\omega_0\right)$.  We will now define other important
constants connected with the wave packets carrier frequency: linear
wave vector $k_0\equiv
k(\omega_0)=\omega_0\sqrt{\varepsilon(\omega_0)}/c$; linear
refractive index $n(\omega_0)=\sqrt{\varepsilon(\omega_0)}$;
nonlinear refractive index
$n_2(\omega_0)=3\pi\chi^{(3)}(\omega_0)/\varepsilon(\omega_0)$;
group velocity:

\begin{eqnarray}
\label{vgr}
v(\omega_0)=\frac{1}{k'}=\frac{c}{\sqrt{\varepsilon(\omega_0)}+
\frac{\omega_0}{2}\sqrt{\frac{1}{\varepsilon}}\frac{\partial\varepsilon}{\partial\omega}},
\end{eqnarray}
nonlinear addition to the group velocity $(\hat{k}_{0nl}^2)'$:

\begin{eqnarray}
\label{vgrn} \left(\hat{k}^2_{0nl}\right)'=\frac{2k_0n_2}{v}+k_0^2
\frac{\partial n_2}{\partial\omega},
\end{eqnarray}
and dispersion of the group velocity $k"(\omega_0)=\partial^2
k/\partial\omega^2_{\omega=\omega_0}.$ All these quantities allow a
direct physical interpretation and we will therefore rewrite
equation (\ref{sve}) in a form consistent with these constants:

\begin{eqnarray}
\label{svea}  -i\left[\frac{\partial\vec A}{\partial t}+
v\frac{\partial\vec A}{\partial
z}+\left(n_2+\frac{k_0v}{2}\frac{\partial n_2}{\partial
\omega}\right)\frac{\partial\left(\left|\vec A\right|^2 \vec
A\right)}{\partial t}\right]
=\nonumber\\
\frac{v}{2k_0}\Delta\vec A
-\frac{v}{2}\left(k"+\frac{1}{k_0v^2}\right)\frac{\partial^2\vec
A}{\partial t^2}+\frac{k_{0}vn_2}{2}\left|\vec A\right|^2 \vec A.
\end{eqnarray}
This equation can be considered to be SVEA of second approximation
with respect to the linear dispersion and of first approximation to
the nonlinear dispersion (nonlinear addition to the group velocity).
It includes the effects of translation in z direction with group
velocity $v$, self-steepening, diffraction, dispersion of second
order and self-action terms. The equations (\ref{svea}) and
(\ref{sve}), with and without the self-steepening term, are derived
in many books and papers \cite{KARP, JAIN, CHRIS, AKH, BOYD}. From
equations (\ref{svea}), (\ref{sve}), after neglecting some of the
differential terms and using a special "moving in time" coordinate
system, it is not hard to obtain the well known spatio-temporal
model. Our intention in this paper is another: before canceling some
of the differential terms in SVEA (\ref{svea}), we must write
(\ref{svea}) in dimensionless form. Then we can estimate and neglect
the small terms, depending on the media parameters, the carrier
frequency and wave vector, and also on the different initial shape
of the pulses. This approach we will apply in Section $5$. As a
result we will obtain equations quite different from the
spatio-temporal ones in the femtosecond region.

\section{Propagation of optical pulses in vacuum and dispersionless media}
The theory of light envelopes is not restricted only to the cases of
non-stationary optical (and magnetic) response. Even in vacuum,
where $\varepsilon=1$  and $\vec P_{nl}=0$, we can write an
amplitude equation by applying solutions of the kind (\ref{a1}) to
the wave equation (\ref{eq8}). We denote here by $\vec V(x,y,z,t)$
the amplitude function  for the electrical field $\vec E(r,t)$ in
vacuum:

\begin{eqnarray}
\label{a1}
 \vec E\left(x,y,z,t\right)=\vec
{V}\left(x,y,z,t\right)\exp{\left(i(k_0z-\omega_0t)\right)},
\end{eqnarray}
where $\omega_0$ and $k_0 $ again are the carrier frequency and the
carrier wave number of the wave packet. We thus obtain the following
linear equation for the amplitude envelope of the electrical field:

\begin{eqnarray}
\label{vac}  -i\left(\frac{\partial\vec V}{\partial t}+
c\frac{\partial\vec V}{\partial z}\right) = \frac{c}{2k_0}\Delta\vec
V -\frac{1}{2k_0c}\frac{\partial^2\vec V}{\partial t^2}.
\end{eqnarray}
The vacuum linear amplitude equation (VLAE) (\ref{vac}) is obtained
directly from the wave equation without any restrictions. This is in
contrast to the case of dispersive medium, where we use the series
of the square of the wave vector and we require the series
(\ref{a9}) to be strongly convergent. That is why equation
(\ref{vac}) describes both amplitudes with many harmonics under the
pulse, and amplitudes with only one or a few harmonics under the
envelope. It is obvious that the envelope $\vec V$ in equation
(\ref{vac}) will propagate with the speed of light $c$ in vacuum.
Equation (\ref{vac}) is valid also for transparent media with
stationary optical response $\varepsilon=const$. In this case, the
propagating constant will be $v=c/\sqrt{\varepsilon\mu}$.

\section {SVEA and VLAE in a normalized form}

Starting from Maxwell's equations for media with non-stationary
linear and nonlinear response, we obtained an amplitude equation and
a SVEA using only two restrictions, which are physically acceptable
for ultra-short pulses. Having adopted the first restriction,
namely, investigation of localized in time and space amplitude
functions only, we introduced the amplitude equation (\ref{ampk}).
Following the second restriction, i.e., limiting ourselves with the
case of a large number of harmonics under the localized envelopes,
we obtained the SVEA (\ref{svea}). As it was pointed out in the
previous section, the second restriction do not affect the VLAE
(\ref{vac}). The next step is writing SVEA (\ref{svea}) and VLAE
(\ref{vac}) in dimensionless variables and estimating the influence
of the different differential terms. In this case, the coefficients
in front of the differential operators in (\ref{svea}) and
(\ref{vac}) will be numbers of different orders, depending on the
medium $n$ and $n_2$, the spectral region of propagation $k_0$ and
$\omega_0$, the field intensity $\left|A_0\right|^2$, and the
initial shape of the pulses, namely, light filament $r_\bot<<z_0$
(LF), Light Bullets (LB)  $r_\bot\approx z_0$ or Light disks (LD)
$r_\bot>>z_0$. With $r_\bot$ we denote here the initial transverse
dimension, "the spot" of the pulse, and with $z_0$ we denote the
initial longitudinal dimension, which is simply the spatial analog
of the initial time duration $t_0$, determined by the relation
$z_0=vt_0$ or $z_0=ct_0$ in the vacuum case. The SVEA (\ref{svea})
and VLAE (\ref{vac}) are written in a Cartesian laboratory
coordinate system. To investigate the dynamics of optical pulses at
long distances, it is convenient to rewrite these equations in a
Galilean coordinate system, where the new reference frame moves with
the group velocity for equation (\ref{svea}), $t' = t; z' = z - vt$:

\begin{eqnarray}
\label{gal} -i\left(\frac{\partial\vec A}{\partial t'}+
\left(n_2+\frac { k_0v}{2}\frac{\partial
n_2}{\partial\omega}\right)\left(\frac{\partial \left(\left|\vec
A\right|^2\vec A\right)}{\partial t'}-\frac{\partial
\left(\left|\vec A\right|^2\vec A\right)}{\partial z'}
\right)\right)=
\frac{v}{2k_0}\Delta_{\bot}\vec A- \nonumber\\
\\
\frac{v^3k_0^{"}}{2}\frac{\partial^2\vec A}{\partial z'^2}
-\frac{v}{2}\left(k"+\frac{1}{k_0v^2}\right)\left(\frac{\partial^2\vec
A}{\partial t'^2}-2v\frac{\partial^2\vec A}{\partial t'\partial
z'}\right)+ \frac{n_2 k_0 v}{2}\left|\vec A\right|^2\vec A,\nonumber
\end{eqnarray}
and with the velocity of light for equation (\ref{vac}), $t' = t; z'
= z - ct$:

\begin{eqnarray}
\label{galvac} -i\frac{\partial\vec V}{\partial t'}=
\frac{c}{2k_0}\Delta_{\bot}\vec V-
\frac{1}{2k_0c}\frac{\partial^2\vec V}{\partial
t'^2}+\frac{1}{k_0}\frac{\partial^2\vec V}{\partial t'\partial z'}.
\end{eqnarray}
With $\Delta_{\bot}=\frac{\partial^2}{\partial x^2} + \frac{\partial
^2}{\partial y^2}$ we denote the transverse Laplacian. We define the
following dimensionless variables connected with the initial
amplitude and with the spatial and temporal dimensions of the pulses
through the relations:

\begin{eqnarray}
\label{eq12} \vec A=A_0\vec A";\ \vec V=V_0\vec V";\ x=r_{\bot} x";\
y=r_{\bot}y";\ z'=z_0z";\ \nonumber \\ t'=t_0t";\ z=z_0z";\ t=t_0t".
\end{eqnarray}
After the substitution of these variables in (\ref{svea}),
(\ref{gal}), (\ref{vac}) and (\ref{galvac}) and making use of the
expressions for the diffraction $ z_{dif}=k_0r_\bot^2$ and
dispersion $z_{disp}=t_0^2/k"$ lengths, we obtain the following five
dimensionless parameters in front of the differential terms in the
equations (\ref{svea}), (\ref{gal}) (\ref{vac}) and (\ref{galvac}):

\begin{eqnarray}
\label{norm}
 \alpha=k_0z_0;\ \delta^2=\frac{r_{\bot}^2}{z_0^2};\ \beta=\frac {z_{dif}}{z_{disp}};\
\gamma=k_0^2r_0^2n_2\left|A_0\right|^2;\nonumber\\
\gamma_1=\left|A_0\right|^2\left(n_2+\frac { k_0v}{2}\frac{\partial
n_2}{\partial\omega}\right).
\end{eqnarray}
Omitting the seconds in the new dimensionless variables and
constants, the equations (\ref{svea}), (\ref{gal}), (\ref{vac}) and
(\ref{galvac}) can be represented as follows:

Case a. SVEA (\ref{svea}) in a laboratory frame ("Laboratory")
\begin{eqnarray}
\label{E13}
 -2i\alpha\delta^2\left(\frac{\partial\vec A}{\partial t}+
\frac{\partial\vec A}{\partial z}+\gamma_1\frac{\partial
\left(\left|\vec A\right|^2\vec A\right)}{\partial t}\right)=
\Delta_{\bot}\vec A +\delta^2\left(\frac{\partial^2\vec A}{\partial
z^2}-\frac{\partial^2\vec A}{\partial t^2}\right)
-\nonumber\\
\\
 \beta\frac{\partial^2\vec A}{\partial
t^2} + \gamma \left|\vec A\right|^2\vec A. \nonumber
\end{eqnarray}

Case b. SVEA (\ref{gal}) in a frame moving with the group velocity:

\begin{eqnarray}
\label{G13}
 -2i\alpha\delta^2\left(\frac{\partial\vec A}{\partial t'}+
\gamma_1\left(\frac{\partial \left(\left|\vec A\right|^2\vec
A\right)}{\partial t'}-\frac{\partial \left(\left|\vec
A\right|^2\vec A\right)}{\partial z'} \right)\right)=
\Delta_{\bot}\vec A -\beta\frac{\partial^2\vec A}{\partial z'^2}
-\nonumber\\
\\
 \left(\beta +\delta^2\right)\left(\frac{\partial^2\vec A}{\partial
t'^2}-2\frac{\partial^2\vec A}{\partial t'\partial z'}\right) +
\gamma \left|\vec A\right|^2\vec A, \nonumber
\end{eqnarray}

Case c. VLAE (\ref{vac}) in a laboratory frame:
\begin{eqnarray}
\label{V13}  -2i\alpha\delta^2\left(\frac{\partial\vec V}{\partial
t}+ \frac{\partial\vec V}{\partial z}\right) = \Delta_{\bot}\vec V
+\delta^2\left(\frac{\partial^2\vec V}{\partial
z^2}-\frac{\partial^2\vec V}{\partial t^2}\right).
\end{eqnarray}

Case d. VLAE (\ref{galvac}) in a Galilean frame:
\begin{eqnarray}
\label{VG13} -2i\alpha\delta^2\frac{\partial\vec V}{\partial t'}=
\Delta_{\bot}\vec V- \delta^2\left(\frac{\partial^2\vec V}{\partial
t'^2}-\frac{\partial^2\vec V}{\partial t'\partial z'}\right).
\end{eqnarray}
It should be noted here that equal dimensionless constants in front
of the differential terms in both the "Laboratory" and "Galilean"
frames are obtained. This gives us the possibility to investigate
and estimate simultaneously the different terms in the normalized
equations (\ref{E13}), (\ref{G13}), (\ref{V13}) and (\ref{VG13}). We
will now discuss these constants in detail, as they play a
significant role in determining the different pulse propagation
regimes.

- The first constant $\alpha=k_0z_0=2\pi z_0/\lambda_0$ determines
with precision $2\pi$ the "number of harmonics" on a FWHM level of
the pulses. Since we use the slowly varying amplitude approximation,
$\alpha$ is always a large number ($\alpha>>1$).

- The second constant $\delta^2=r_\bot^2/z_0^2$ determines the
relation between the initial transverse and longitudinal size of the
optical pulses. This parameter distinguishes the case of light
filaments (LF) $\delta^2=r_\bot^2/z_0^2<<1$ from the case of LB
$\delta^2=r_\bot^2/z_0^2\cong1$ and the case of light disks LD
$\delta^2=r_\bot^2/z_0^2>>1$. For light filaments $\delta^2<<1$ and
we can neglect the differential terms with coefficient $\delta^2$.
It is not  difficult to see that in this case the SVEA (\ref{E13}),
(\ref{G13}) and VLAE (\ref{G13}), (\ref{V13}) can be transformed to
the standard paraxial approximation of the linear and nonlinear
optics. If we set the possible values of the optical pulses'
transverse dimensions at $3-4 mm > r_{\bot} >100 \mu m$, we can
directly obtain the above distinction in dependence on the time
duration of the pulses. For light pulses with time duration $ns >t_0
> 40-50 ps$ we obtain $\delta^2=r_\bot^2/z_0^2<<1$ and we are in the
regime of LF and paraxial approximation. In the case of light pulses
with duration $t_0\approx 3-4 ps$ up to $500-600 fs$ it is possible
to reach $\delta^2=r_\bot^2/z_0^2\cong1$ and we are in the regime of
LB. For pulses in the time range $300 fs-30 fs$ we can prepare the
initial shape of the pulses to satisfy the relation
$\delta^2=r_\bot^2/z_0^2>>1$ and thus reach the LD regime. It is
important to note here that the wave packets in the visible and UV
ranges with time duration $t_0\geq 30 fs$ contain more than 10-15
optical harmonics under the pulse, so that we are still in SVEA
approximation. In the last two cases (LB and LD) the differential
terms with $\delta^2$ cannot be ignored and the equations
(\ref{E13}), (\ref{G13}) and (\ref{V13}) governing the propagation
of pulses with initial form of LB and LD are quite different from
the paraxial approximation.

-The third parameter is $\beta=k_0r_\bot^2/z_{dis}$, where
$z_{dis}=t_0^2/k"$ determines the relation between the diffraction
and dispersion lengths. The dispersion parameter $k"$ in the visible
and UV transparency region of dielectrics has values from $k"\sim
10^{-31}$ $s^2/cm$  for gases and metal vapors up to $k"\sim
10^{-26}$ $s^2/cm$ for solid materials. It is convenient to express
this parameter using the product of the second constant $\delta^2$
and the parameter $\beta_1=k_0v^2k"$ by the relation
$\beta=\beta_1\delta^2$. For typical values of the dispersion $k"$
in the visible and UV region listed above, the dimensionless
parameter $\beta_1$ is very small ($\beta_1<<1$), while for optical
pulses propagating in the UV region in solids and liquids it may
reach $\beta_1\propto 1$. The parameter $\beta_1$ can be also
negative and may reach $\beta_1\simeq -1$ near electronic resonances
and also near the Langmuir frequency in electronic plasmas.  As it
is shown in \cite{KOV}, only in the case $\beta_1\simeq -1$  we can
obtain the 3D+1 nonlinear Schrodinger equation from SVEA (\ref{E13})
and (\ref{G13}).

- The fourth and fifth constants
$\gamma=k_0^2r_{\perp}^2n_2\left|A_0\right|^2$ and
$\alpha\delta^2\gamma_1\simeq\alpha\delta^2n_2\left|A_0\right|^2$
are correspondingly the nonlinear coefficient and the coefficient of
nonlinear addition to the group velocity (coefficient before the
front of the first order nonlinear dispersion term). It is easy to
estimate that for $\alpha>>1$ and $\delta^2\geq1$, we always have
$\gamma>>\alpha\delta^2\gamma_1$. For optical pulses with power near
the critical threshold for self-focusing $\gamma\cong1$ and less
(linear regime) $\gamma<<1$, the nonlinear addition to the group
velocity is very small ($\alpha\delta^2\gamma_1<<1$) and from here
to the end of this paper we will neglect the terms with the first
addition to the nonlinear dispersion. The analysis of the
dimensionless constants performed above leads us to the following
conclusion: Dynamics of wave packets with power near to critical for
self-focusing $\gamma\propto1$ in the visible and UV region in a
media with dispersion are governed by the following SVEA equations:

Case a. SVEA in laboratory frame ("Laboratory")
\begin{eqnarray}
\label{LAB13}
 -2i\alpha\delta^2\left(\frac{\partial\vec A}{\partial t}+
\frac{\partial\vec A}{\partial z}\right)= \Delta_{\bot}\vec A
+\delta^2\frac{\partial^2\vec A}{\partial z^2}
 -\delta^2\left(\beta_1+1\right)\frac{\partial^2\vec A}{\partial
t^2} + \gamma \left|\vec A\right|^2\vec A.
\end{eqnarray}
Case b. SVEA in frame moving with group velocity ("Galilean"):

\begin{eqnarray}
\label{GAL13}
 -2i\alpha\delta^2\frac{\partial\vec A}{\partial t'}=
\Delta_{\bot}\vec A -\beta_1\delta^2\frac{\partial^2\vec A}{\partial
z'^2}-\nonumber\\ \delta^2\left(\beta_1
+1\right)\left(\frac{\partial^2\vec A}{\partial
t'^2}-2\frac{\partial^2\vec A}{\partial t'\partial z'}\right) +
\gamma \left|\vec A\right|^2\vec A.
\end{eqnarray}
Equations (\ref{LAB13}) and (\ref{GAL13}) are quite different from
the well known paraxial spatio-temporal evolution equations. Here
are included also the second derivative along the $z$ direction, a
mixed term and additional second derivative in time term. This leads
to  dynamics of the ultrashort fs pulses different from
spatio-temporal model. In this paper we will investigate the
propagation in linear regime, when $\gamma<<1$.

\section{Fundamental solutions of the linear SVEA and VLAE}

The behavior of long pulses is similar to that of optical beams,
since their propagation is governed by a equation where the
nonparaxial terms become small. That is why we can expect the
diffraction enlargement of long pulses to be of the same order as
are the optical beams. The situation regarding LB and LD is
different. Their propagation is governed by equations in media with
non-stationary optical response - SVEA (\ref{LAB13}) and
(\ref{GAL13}), and by VLAE (\ref{vac}) and (\ref{galvac}) in media
with linear stationary optical response (or vacuum), where the
nonparaxial terms are of same order and bigger  than  transverse
Laplacian. In this section we will solve the equations
(\ref{LAB13}), (\ref{GAL13}) in linear regime ($\gamma<<1$) and will
compare the solutions with the solutions of the linear VLAE
(\ref{vac}) and (\ref{galvac}). Neglecting the small nonlinear terms
in (\ref{LAB13}), (\ref{GAL13}) we obtain:

a. Linear SVEA in a laboratory coordinate frame:

\begin{eqnarray}
\label{eqlin13}
 -2i\alpha\delta^2\left(\frac{\partial\vec A}{\partial t}+
\frac{\partial\vec A}{\partial z}\right)= \Delta_{\bot}\vec A
+\delta^2\frac{\partial^2\vec A}{\partial
z^2}-\delta^2\left(\beta_1+1\right)\frac{\partial^2\vec A}{\partial
t^2}.
\end{eqnarray}
b. Linear SVEA in a Galilean coordinate frame:

\begin{eqnarray}
\label{lin13}
 -2i\alpha\delta^2\frac{\partial\vec A}{\partial t'}=
\Delta_{\bot}\vec A
-\delta^2\left(\beta_1+1\right)\left(\frac{\partial^2\vec
A}{\partial t'^2}-2\frac{\partial^2\vec A}{\partial t'\partial
z'}\right)-\delta^2\beta_1\frac{\partial^2\vec A}{\partial z'^2}.
\end{eqnarray}
For comparison we will rewrite here the corresponding linear VLAE:

c. Linear VLAE  in a laboratory frame:

\begin{eqnarray}
\label{VV13}  -2i\alpha\delta^2\left(\frac{\partial\vec V}{\partial
t}+ \frac{\partial\vec V}{\partial z}\right) = \Delta_{\bot}\vec V
+\delta^2\left(\frac{\partial^2\vec V}{\partial
z^2}-\frac{\partial^2\vec V}{\partial t^2}\right).
\end{eqnarray}
d. Linear VLAE in a Galilean frame:

\begin{eqnarray}
\label{VVG13} -2i\alpha\delta^2\frac{\partial\vec V}{\partial t'}=
\Delta_{\bot}\vec V- \delta^2\left(\frac{\partial^2\vec V}{\partial
t'^2}-\frac{\partial^2\vec V}{\partial t'\partial z'}\right).
\end{eqnarray}

As can be expected, the equations for ultra-short optical pulses in
vacuum and dispersionless media (\ref{VV13}), (\ref{VVG13}) become
identical with the equations with dispersion (\ref{eqlin13}),
(\ref{lin13}) when the dispersion parameter $ \beta_1$ and
$\delta^2\beta_1$ are small. In equations (\ref{eqlin13}),
(\ref{lin13}) there are three dimensionless parameters, $\alpha$,
$\delta^2$ and $\beta_1$ ,while in (\ref{VV13}), (\ref{VVG13}) there
are only two $\alpha$ and $\delta^2$. These parameters can be
changed considerably in fs region, and this leads, as we can see
later, to quite different dynamics in the particular cases.

In the general case we apply the Fourier method to solve the linear
SVEA (\ref{eqlin13}), (\ref{lin13}) which describe the propagation
of ultrashort optical pulses in medium with dispersion and linear
VLAE (\ref{VV13}), (\ref{VVG13}) which govern the propagation of
light pulses in vacuum and dispersionless media. We mark the Fourier
transform of the amplitude functions of SVEA  in Galilean frame
(\ref{eqlin13}) with $\vec A_G(k_x,k_y,k_z,t)=F(\vec{A}(x,y,z,t))$,
and in  Lab frame (\ref{lin13}) with $\vec
A_L(k_x,k_y,k_z,t)=F(\vec{A}(x,y,z,t))$. The Fourier transform of
the amplitude functions of VLAE  in  Galilean frame become
(\ref{VV13}), $\vec B_G(k_x,k_y,k_z,t)=F(\vec{V}(x,y,z,t))$, while
in  Laboratory frame (\ref{VVG13}) we write $\vec
B_L(k_x,k_y,k_z,t)=F(\vec{V}(x,y,z,t))$. Applying spatial Fourier
transformation to the components of the amplitude vector functions
$\vec A$ and $\vec V$, the following ordinary linear differential
equations in $k_x,k_y,k_z$ space for SVEA:

a. Laboratory:

\begin{eqnarray}
\label{Fe13}
 -2i\alpha\delta^2\frac{\partial\vec A_L}{\partial t}=\nonumber\\
-\left({k_x}^2+{k_y}^2+\delta^2({k_z}^2-2\alpha k_z)\right)\vec
A_L-\delta^2\left(\beta_1+1\right)\frac{\partial^2\vec A_L}{\partial
t^2},
\end{eqnarray}
b. Galilean:

\begin{eqnarray}
\label{Fl13}
-2i\delta^2\left(\alpha-\left(\beta_1+1\right)k_z\right)\frac{\partial\vec
A_G}{\partial t}=\nonumber\\
-\left({k_x}^2+{k_y}^2-\delta^2\beta_1k_z^2\right)\vec
A_G-\delta^2\left(\beta_1+1\right)\frac{\partial^2\vec A_G}{\partial
t^2},
\end{eqnarray}
and the following equations for VLAE:

c. Laboratory:

\begin{eqnarray}
\label{FV13} -2i\alpha\delta^2\frac{\partial\vec B_L}{\partial t}=-
\left({k_x}^2+{k_y}^2+\delta^2({k_z}^2-2\alpha k_z)\right)\vec
B_L-\delta^2\frac{\partial^2\vec B_L}{\partial t^2},
\end{eqnarray}
d. Galilean:

\begin{eqnarray}
\label{FGV13}
 -2i\delta^2\left(\alpha-k_z\right)\frac{\partial\vec B_G}{\partial
 t}=
-\left({k_x}^2+{k_y}^2\right)\vec B_G-\delta^2\frac{\partial^2\vec
B_G}{\partial t^2},
\end{eqnarray}
are obtained. We look for solutions of the kind  of $\vec
A_L=\vec{A}_L(k_x,k_y,k_z)\exp(i\Omega_L t)$ and $\vec
A_G=\vec{A}_G(k_x,k_y,k_z)\exp(i\Omega_G t)$ for the equations
(\ref{Fe13}), (\ref{Fl13}), and for solutions  $\vec
B_L=\vec{B}_L(k_x,k_y,k_z)\exp(i\Phi_L t)$ and $\vec
B_G=\vec{B}_G(k_x,k_y,k_z)\exp(i\Phi_G t)$ for the equations
(\ref{FV13}) and (\ref{FGV13}) correspondingly. Let us denote the
square of the sum of the wave vectors  as:
$\hat{k}^2={k_x}^2+{k_y}^2+\delta^2({k_z}^2-2\alpha k_z)$. The
solutions exist when $\Omega_L$, $\Omega_G$,  $\Phi_L$ and $\Phi_G$
satisfy the following quadratic equations:

\begin{eqnarray}
\label{L}
\Omega_L^2-2\frac{\alpha}{\beta_1+1}\Omega_L-\frac{\hat{k}^2}{\delta^2(\beta_1+1)}=0,
\end{eqnarray}
\begin{eqnarray}
\label{G}
\Omega_G^2-2\frac{\left(\alpha-(\beta_1+1)k_z\right)}{\beta_1+1}\Omega_G-
\frac{{k_x}^2+{k_y}^2-\delta^2\beta_1k_z^2}{\delta^2(\beta_1+1)}=0.
\end{eqnarray}

\begin{eqnarray}
\label{VL} \Phi_L^2-2\alpha\Phi_L-\frac{\hat{k}^2}{\delta^2}=0,
\end{eqnarray}
\begin{eqnarray}
\label{VG}
\Phi_G^2-2(\alpha-k_z)\Phi_G-\frac{{k_x}^2+{k_y}^2}{\delta^2}=0.
\end{eqnarray}
The solutions of (\ref{L}), (\ref{G}) for media with dispersion are:

\begin{eqnarray}
\label{OmegaL}
{\Omega_L}^{1,2}=\frac{\alpha}{\beta_1+1}\pm\sqrt{\frac{\alpha^2}{(\beta_1+1)^2}+\frac{\hat{k}^2}{\delta^2(\beta_1+1)}},
\end{eqnarray}
\begin{eqnarray}
\label{OmegaG}
{\Omega_G}^{1,2}=\frac{\alpha-(\beta_1+1)k_z}{\beta_1+1}\pm
\sqrt{\frac{\left(\alpha-(\beta_1+1)k_z\right)^2}{(\beta_1+1)^2}+\frac{k_x^2+k_y^2-
\delta^2\beta_1k_z^2}{\delta^2(\beta_1+1)}},
\end{eqnarray}

while the solutions of (\ref{VL}), and (\ref{VG}) for dispersionless
media and vacuum become:
\begin{eqnarray}
\label{PhiL}
{\Phi_L}^{1,2}=\alpha\pm\sqrt{\alpha^2+\hat{k}^2/\delta^2},
\end{eqnarray}
\begin{eqnarray}
\label{PhiG}
{\Phi_G}^{1,2}=(\alpha-k_z)\pm\sqrt{\alpha^2+\hat{k}^2/\delta^2}.
\end{eqnarray}
Now the necessity of a parallel investigation of the propagation of
optical pulses in media with dispersion, in dispersionless media and
in vacuum becomes obvious. Further, we will introduce here the
concept of weak and strong dispersion media depending on the value
of dimensionless dispersion parameter $\beta_1$. When $\beta_1<<1$
we have media with weak dispersion. It is not difficult to calculate
that in the optical transparency region of gases, liquids and solid
materials, this parameter is usually very small ( $\beta_1<<1$). For
media with weak dispersion the solutions of the characteristic
equations with dispersion (\ref{OmegaL}), (\ref{OmegaG}) are
identical to the solutions for media without dispersion and vacuum
(\ref{PhiL}), (\ref{PhiG}) correspondingly. In media with strong
dispersion $\beta_1$ can reach the values $\beta_1\simeq1-3$ in the
UV transparency region of solids and liquids. In this case, the
solutions for media with dispersion will be slightly modified with
the factor $\beta_1+1$ with respect to the solutions without
dispersion. We consider here the regime of propagation far away from
electronic resonances and the Langmuir frequency in electronic
plasmas, where it is possible to obtain a strongly negative
dispersion parameter $\beta_1\propto-1$. We point out here again
that in the case of LB, when $\beta_1\cong-1$, the amplitude
equations (\ref{svea}) can be transformed into the 3D+1 linear and
nonlinear vector Schrodinger equations \cite{KOV}. Generally said,
the dispersion parameter $\beta_1$ varies slowly from the visible to
the UV transparency region of the materials from very small values
up to $\beta_1\simeq 1-3$ and that is why it does not influence
radically the solutions and the propagation of optical pulses in
linear regime. The other parameters $\alpha$ and $\delta^2$ change
significantly. For example $\alpha$ varies from $10^1$ to $10^3$,
while $\delta^2$ varies from $10^{-2}-10^{-4}$ for LF to $10^0$ for
LB and $10^2-10^4$ for LD. This is the reason to investigate more
precisely in the next paragraph the solutions of the equations for
media with weak dispersion as air where $\beta_1<<1$,
(\ref{OmegaG}), as we expect that the solutions for media with
strong dispersion (UV transparency region of solids and liquids)
will be only slightly modified by the factor $\beta_1+1;\
\beta_1\leq 1 $. We obtained solutions of the characteristic
equations (\ref{OmegaL}), (\ref{OmegaG}), (\ref{PhiL}) and
(\ref{PhiG}). The solutions of the corresponding linear differential
equations  SVEA (\ref{Fe13}), (\ref{Fl13}) and VLAE (\ref{FV13}),
(\ref{FGV13}) in the $k$-space become:

a. Solution of SVEA in the k-space and laboratory coordinate frame:

\begin{eqnarray}
\label{SOL} \vec{A_L}=\vec{A_L}(k_x,k_y,k_z,t=0)\times\nonumber\\
\exp\left(i\left(\frac{\alpha}{\beta_1+1}
\pm\sqrt{\frac{\alpha^2}{(\beta_1+1)^2}+\frac{\hat{k}^2}{\delta^2(\beta_1+1)}}\right)t\right).
\end{eqnarray}
b. Solution of SVEA in the k-space and  Galilean coordinate frame:

\begin{eqnarray}
\label{SOG}\vec{A_G}=\vec{A_G}(k_x,k_y,k_z,t=0)\times\nonumber\\
\\
\exp\left(i\left(\frac{\alpha-(\beta_1+1)k_z}{\beta_1+1}
\pm\sqrt{\frac{\left(\alpha-(\beta_1+1)k_z\right)^2}{(\beta_1+1)^2}+\frac{k_x^2+k_y^2-
\delta^2\beta_1k_z^2}{\delta^2(\beta_1+1)}}\right)t\right)\nonumber.
\end{eqnarray}
c. Solution of VLAE in k-space and laboratory coordinate frame:

\begin{eqnarray}
\label{BSOL}
\vec{B_L}=\vec{B_L}(k_x,k_y,k_z,t=0)\exp\left(i\left(\alpha\pm\sqrt{\alpha^2+\hat{k}^2/\delta^2}\right)t\right).
\end{eqnarray}
d. Solution of VLAE in the k-space and  Galilean coordinate frame:

\begin{eqnarray}
\label{BSOG}\vec{B_G}=\vec{B_G}(k_x,k_y,k_z,t=0)\exp
\left(i\left((\alpha-k_z)\pm\sqrt{\alpha^2+\hat{k}^2/\delta^2}\right)t\right).
\end{eqnarray}
It is obvious that the solutions (\ref{BSOL}) and (\ref{BSOG}) of
equations (\ref{FV13}) and (\ref{FGV13}) should be equal with
accuracy - a wave number in z direction. This follows from the
Fourier transform of such evolution equations and leads to only one
difference between the solutions in the real space - the motion of
the pulse in the z direction in a Laboratory frame and its
stationarity in a Galilean frame. As it was pointed out at the
beginning, we investigate here only localized in space and time
initial functions of the amplitude envelopes. Thus, the images of
these functions after Fourier transform in the $k_x,k_y,k_z$ space
are also localized functions. The solutions of our amplitude
equations in $k$ space (\ref{SOL}), (\ref{SOG}),(\ref{BSOL}) and
(\ref{BSOG}) are the product of the initial localized in
$k_x,k_y,k_z$-space functions and the new spectral kernels which are
periodic (different from the Fresnel's one). The product of a
localized function and a periodic function is also a function
localized in the $k_x,k_y,k_z$ space. Therefore, the solutions of
our amplitude equations in $k$ space (\ref{BSOL}),
(\ref{BSOG}),(\ref{SOL}) and (\ref{SOG}) are also localized
functions in this space and we can apply the inverse Fourier
transform to obtain again the fundamental localized solutions in the
$x,y,z,t$ space. More precisely, we use the convolution theorem to
present our fundamental solutions in the real space as a convolution
of inverse Fourier transform of the initial pulse with the inverse
Fourier transforms of the new spectral kernels:

a. Fundamental solution of SVEA (\ref{eqlin13}) in  laboratory
coordinate frame:

\begin{eqnarray}
\label{L1} \vec{A}(x,y,z,t)=F^{-1}\left(\vec{A_L}(k_x,k_y,k_z,t=0)\right)\otimes\nonumber\\
F^{-1}\left(\exp\left(i\left(\frac{\alpha}{\beta_1+1}
\pm\sqrt{\frac{\alpha^2}{(\beta_1+1)^2}+\frac{\hat{k}^2}{\delta^2(\beta_1+1)}}\right)t\right)\right).
\end{eqnarray}
b. Fundamental solution of SVEA (\ref{lin13}) in  Galilean
coordinate frame:

\begin{eqnarray}
\label{LG}\vec{A}(x,y,z',t')=F^{-1}\left(\vec{A_G}(k_x,k_y,k_z,t=0)\right)\otimes\nonumber\\
\\
F^{-1}\left(\exp\left(i\left(\frac{\alpha-(\beta_1+1)k_z}{\beta_1+1}
\pm\sqrt{\frac{\left(\alpha-(\beta_1+1)k_z\right)^2}{(\beta_1+1)^2}+\frac{k_x^2+k_y^2-
\delta^2\beta_1k_z^2}{\delta^2(\beta_1+1)}}\right)t\right)\right).\nonumber
\end{eqnarray}
c. Fundamental solution of VLAE (\ref{VV13}) in  laboratory
coordinate frame:

\begin{eqnarray}
\label{L2}
\vec{V}(x,y,z,t)=F^{-1}\left(\vec{B_L}(k_x,k_y,k_z,t=0)\right)\otimes\nonumber\\
F^{-1}\left(\exp\left(i\left(\alpha\pm\sqrt{\alpha^2+\hat{k}^2/\delta^2}\right)t\right)\right).
\end{eqnarray}
d. Fundamental solution of VLAE (\ref{VVG13}) in  Galilean
coordinate frame:

\begin{eqnarray}
\label{LG2}
\vec{V}(x,y,z',t')=F^{-1}\left(\vec{B_G}(k_x,k_y,k_z,t=0)\right)\otimes\nonumber\\
F^{-1}\left(\exp
\left(i\left((\alpha-k_z)\pm\sqrt{\alpha^2+\hat{k}^2/\delta^2}\right)t\right)\right),
\end{eqnarray}
where with $F^{-1}$ we denote the spatial three-dimensional inverse
Fourier transform and with $ \otimes$ we denote the convolution
symbol. The difference between the Fresnel's integrals, describing
propagation of optical beams and long pulses in linear regime, and
the new integrals (\ref{L1}), (\ref{LG}), (\ref{L2}) and
(\ref{LG2}), which are  solutions of the linear evolution equations
(\ref{eqlin13}), (\ref{lin13}), (\ref{VV13}) and (\ref{VVG13}) is
quite obvious. In addition, in the new spectral kernels there are
three dimensionless parameters: $\alpha$, $\delta^2$ and $\beta_1$.
Let us fix $\alpha$ to be always large, i.e. $\alpha>>1$. As pointed
out above, the condition $\alpha>>1$ is not necessary for VLAE, so
we can investigate pulses with longitudinal duration of the order of
the carrier wavelength in vacuum and in dispersionless media. To
provide one qualitative  analyze for the influence of the other two
parameters, $\delta^2$ and $\beta_1$,  on the evolution of the
initial pulse, we will rewrite the expression for the spectral
kernel (\ref{OmegaL}) of the solutions (\ref{L1}) of equation
(\ref{eqlin13}) in the following form:

\begin{eqnarray}
\label{OmL}
{\Omega_L}^{1,2}=\frac{\alpha}{\beta_1+1}\nonumber\\
\pm\sqrt{\frac{\alpha^2}{(\beta_1+1)^2}+\frac{1}{\delta^2(\beta_1+1)}(k_x^2+k_y^2)+
\frac{1}{\beta_1+1}(k_z^2-2\alpha k_z)}.
\end{eqnarray}

As $\alpha>>1$  and $ \beta_1\leq 1 $, the diffraction widening will
be determined by the second term under the square root in
(\ref{OmL}):
\begin{eqnarray}
\label{kern} \frac{1}{\delta^2(\beta_1+1)}(k_x^2+k_y^2),
\end{eqnarray}
which determines the transverse  diffraction and dispersion widening
of the pulses. We pointed out above that the dispersion parameter
varies very slowly within the limits $0\leq\beta_1<10^1$, while the
relations between the transverse and longitudinal part varies
significantly $10^{-4}<\delta^2<10^4$. This is why we  estimate
mainly the influence of the different values of $\delta^2$ on the
diffraction widening. We investigate the following basic cases:

a/ Long pulses, when $\delta^2<<1$. It is easy to estimate from
(\ref{kern}), that the transverse enlargement ${k_x}^2+{k_y}^2$ will
dominate significantly as:
\begin{eqnarray}
\label{KL} \frac{1}{\delta^2(\beta_1+1)}>>1 .
\end{eqnarray}
In the case of long pulses we have also $\alpha\delta^2\sim 1$ and
as it can be seen from equations (\ref{Fe13}) and (\ref{Fl13}) we
have similar (\emph{but not equal}) diffraction length to that of
optical beam $(z^{beam}_{diff}=k_0r^{2}_{\perp})$. The difference is
only in the factors $\alpha\delta^2\sim 1$. When pulse propagate in
optical transparency region of the air and gases $\beta_1<<1$,
normalized dispersion parameter is to small that the main factor
which determinate the diffraction widening become
$z^{pulse}_{diff}=\alpha\delta^2z^{beam}_{diff}=k_0^2r^{4}_{\perp}/z_0$.
The validity of this new diffraction formula for pulses will be
studied more precisely in the next paragraph not only for long
pulses but also for LB and LD.

b/ LB: $\delta^2\simeq 1$. In the case of optical pulses with
approximately equal transverse and longitudinal size, we obtain the
following coefficient of the transverse ${k_x}^2+{k_y}^2$
diffraction terms:
\begin{eqnarray}
\label{KB} \frac{1}{\delta^2(\beta_1+1)}\cong 1/2 .
\end{eqnarray}
Hence, the diffraction and dispersion transverse enlargement will be
reduced by the factor $\delta^2(\beta_1+1)$ with respect to the
diffraction of long pulses and Fresnel's diffraction.

In addition, we will point out here an important asymptotic behavior
of LB: When $\alpha^2$ is small (pulses with only  few harmonics
under the envelope) and $\beta_1<<1$ (media with weak dispersion,
dispersionless media and vacuum), the spectral kernels of the new
equations tend to the asymptotical value $\sim
\exp\left(i(\sqrt{k_x^2+k_y^2+(k_z-\alpha)^2})t\right)\cong\exp\left(i(|k|t)\right)$,
which is actually the spectral kernel of the 3D wave equation. For
this reason we can expect for optical pulses with only one or two
harmonics under the envelope (subfemtosecond and attosecond pulses)
diffraction similar to the typical diffraction of the 3D wave
equation, whose dynamics is characterized by internal and external
fronts and a significant widening of the pulse.

c/ Light disks: This is the case when the longitudinal size $z_0$ is
mush shorter than the transverse size $r_\bot$ and $\delta^2>>1$. As
indicated above, the typical time region for such pulses is $30-40
fs<t_0<200-300 fs$. We determine the lower limits of this relation
from the condition $\alpha^2>>1$, i.e a large number of harmonics
under the envelope. This condition still holds true for pulses in
the visible and UV regions with time duration $30-40 fs$. The
dimensionless parameter in front of the transverse diffraction and
dispersion $\left({k_x}^2+{k_y}^2\right)$ will be of the order of:

\begin{eqnarray}
\label{KLL} \frac{1}{\delta^2(\beta_1+1)}<<1.
\end{eqnarray}
We thus see that the transverse enlargement is of the order of
$\delta^2(\beta_1+1)$, or negligible as compared with LB, and
smaller by a factor of about $(\delta^2(\beta_1+1))^2$ than in the
cases of long pulses and Fresnel's diffraction. To summarize the
results of this section, we can expect that the transverse
diffraction and dispersion enlargement of LB should be smaller by a
factor $\delta^2(\beta_1+1)$ than those of LF, while the transverse
diffraction and dispersion enlargement of LD should be smaller by a
factor of $(\delta^2(\beta_1+1))^2$ than those of long pulses and
paraxial approximation. Practically no transverse enlargements of LD
would be observed over long distances, namely, more than tens  and
hundred of diffraction length.

\begin{figure}[ht]\label{Fig1}
\begin{center}
\includegraphics[width=120mm,height=40mm]{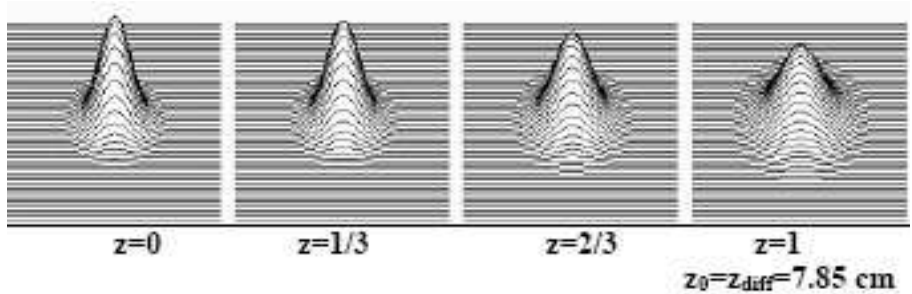}
\caption{Intensity profiles  of a Gaussian beam with initial
condition $A_x(x,y,z=0)=\exp\left(-\frac{x^2+y^2}{2}\right)$
governed by the $2D$ paraxial equation (\ref{plin13}). The
transverse size (the spot) grows  by factor $ \sqrt{2}$ over
normalized distance $z=1$. This correspond to a real distance
$z_0=z_{diff}^{beam}=7.85$ $cm$ for the selected in the paper laser
source on $\lambda_0=800$ $nm$.}
\end{center}
\end{figure}

\section{Dynamics of light beam, LF, LB and LD in air }
In the beginning of this section we will discuss more widely the
Galilean invariancy and connections between normalized equations
written in different coordinate systems: Laboratory and Galilean.
The Galilean invariancy of the SVEA (\ref{eqlin13}) and VLAE
(\ref{VV13}) is not obvious. Indeed, after using the transformation
$t'=t$ and $z'=z-vt$ where $v=1$ the new equations in Galilean
frame, (\ref{lin13}) and (\ref{VVG13}) admit mixed $z,t$ terms and
look quite different. The Galilean invariancy can be seen only from
the kind of the fundamental solutions of  equations
(\ref{L1})-(\ref{LG2}) in Laboratory and Galilean frames. The
solutions are equal with precision wave number $k_z$, which gives
the stationarity in Galilean and translation in z direction in
Laboratory frame. The numerical solutions  with initial conditions -
Gaussian pulses provided in \cite{KOV1} for both coordinate systems
demonstrate again that localized waves admit equal spatial and phase
deformation in both coordinates and there is only one difference -
stationarity in Galilean  and translation with normalized velocity
v=1 in Laboratory frame. Naturally, the best way is to solve
numerically the equations (\ref{eqlin13}) or (\ref{VV13}) in
Laboratory system and to see one spatial and phase transformation of
the pulse as well as its translation. Inconvenience in such one
approach is, that at long distances the pulses will move out of the
grid. And here the Galilean invariancy of the normalized equations
help us. It is well known that the times in Galilean and Laboratory
systems are equal $t'=t$. As the normalized velocity is $v=1$, the
same translation $z=1$ in Laboratory frame correspond to time
evolution in both system $t'=t=1$ . This is demonstrated in
\cite{KOV1} but here it gives us one additional opportunity. We can
solve the equation (\ref{lin13}) in Galilean frame for long time t'
without pulses to move out of the grid and to connect the normalized
time t' in Galilean with normalized time $t$ and translation $z$ in
Laboratory frame $t'=t=z;\ v=1$. After that, using normalized
constants, we can obtain the real distance of propagation of optical
pulses. It is not hard to see that normalized distance $z=1$
correspond  to a real distance $z_0=k_0r_{\perp}^2$ when $
\alpha\delta^2=1$. And this is the natural way to compare the
dynamics of optical pulses governed by equation (\ref{lin13}) in
Galilean frame with the evolution of a laser beam in scalar paraxial
approximation described by normalized equation:

\begin{eqnarray}
\label{plin13}
 -i\frac{\partial A}{\partial z}+\frac{1}{2}\Delta_{\bot} A=0,
\end{eqnarray}
where, as it is well known, $z=1$ in normalized coordinates
corresponds to a real distance $z_0=k_0r_{\perp}^2$ called
diffraction length. This length determines the distance where the
laser beam increase its width on level $e^{-1}$ from the maximum
with factor$\sqrt{2}$. We investigate here only laser sources with
spectrally limited, not phase modulated initial Gaussian profile.
The optical lens and devices add additional phase modulation on the
initial pulse and influenced on the widening in linear regime.

Evolution of real laser pulse with the following characteristics is
considered: light source form Ti:sapphire with width on level
$e^{-1}$; $r_{\perp}=100$ $\mu m$. Usually such small spot of the
pulse is made by focusing by lens. To obtain no modulated in phase
initial pulse the additional phase from the lens must be reduced to
zero by a system of lens. The solutions of linear SVEA (\ref{lin13})
in Galilean frame are carried out for optical wave on wave-length
$\lambda_0=800 $ $nm$ propagating in air and the following
constants: carrying wave number $k_0=n_b\omega/c=7.854\times 10^4$;
$cm^{-1}$, where $n_b\approx 1.0$ for air; GVD coefficient $k"= 3.0
\times 10^{-31}$ $sec^2/cm$; normalized GVD coefficient
$\beta_1=k_0v^2k" \cong 2.1 \times 10^{-4}$. To find the difference
in dynamics of LF, LB, and LD we select different time duration of
pulses for LF ($t_0=260 ps$), LB ($t_0=330 fs)$ and LD $(t_0=33
fs$). Using the above parameters of the laser sources and the
material constants we obtain the following dimmensionless parameters
in SVEA (\ref{lin13}) for the particular cases:

a) long pulse and $\alpha\delta^2=1$ ($t_0=260 ps$):
$\alpha=6.16\times10^{5}$; $\delta^2=1/6.16\times10^{-5}$;
$\beta_1=2.1 \times 10^{-4}$.

b) light bullet (330 fs):  $\alpha=785.0$; $\delta^2=1.0$;
$\beta_1=2.1 \times 10^{-4}$.

c) light disk (33 fs): $\alpha=78,5$; $\delta^2=100.0$; $\beta_1=2.1
\times 10^{-4}$.

Other important parameter for comparing the pulse dynamics and
paraxial evolution of a laser beam is the diffraction length
$z_{diff}= k_0r_{\perp}^2=7.85$ $cm$. In addition, we should point
out that all coming numerical computations are performed with pulses
satisfying the boundary conditions $\lim_{x.y,z\mapsto \pm L/2}
A(x,y,z,t)\vec{x}=0$ and also $\lim_{x.y,z\mapsto \pm \Lambda/2}
A(k_x,k_y,k_z,t)=0$, where $L$ and $\Lambda$ are respectively the
spatial and wave-number intervals for the calculations.

\subsection{Evolution of Gaussian beam in paraxial approximation}
The initial conditions for linearly polarized normalized Gaussian
beam reads:

\begin{eqnarray}
\label{PGAUS}
\vec{A}=A_x\vec{x};\nonumber\\
A_x(x,y,z=0)=\exp\left(-\frac{x^2+y^2}{2}\right).
\end{eqnarray}
The evolution of the initial Gaussian beam (\ref{PGAUS}) governed by
the paraxial equation (\ref{plin13}) is described by the Fresnel's
integral  or can be found by numerical calculation of the inverse
Fourier transform of the solution in the ($k_x,k_y$)-space. The
intensity profile of a solution $A(x,y,z)$ of the paraxial equation
(\ref{plin13}) with initial condition (\ref{PGAUS}) on a normalized
distance $z=1$ is illustrated on Fig.1. Getting in ming the above
real parameters of a laser system on $800$ $nm$, the normalized
distance $z=1$ corresponds to one diffraction length and real
distance of $z_0=z_{diff}= k_0r_{\perp}^2=7.85$ $cm$.

\begin{figure}[ht]\label{Fig2}
\begin{center}
\includegraphics[width=100mm,height=50mm]{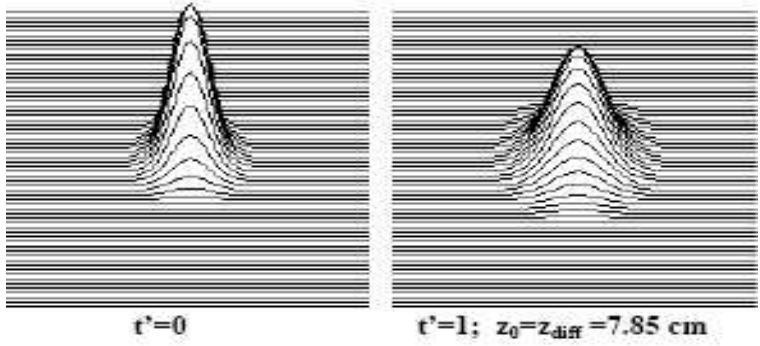}
\caption{Transverse intensity distribution of $260$ $ps$ pulse on
carrying frequency $800$ $nm$ (long Gaussian pulse). Numerical
solutions of the linear SVEA (\ref{lin13}) in Galilean frame is
performed by the following particularly selected initial conditions
to satisfy $\alpha\delta^2=1$:
$A_x(x,y,z,t=0)=\exp\left(-\frac{x^2+y^2+z^2}{2}\right)$,
$\alpha=6.16\times10^{5}$;\ $\delta^2=1/6.16\times10^{-5}$;\
$\beta_1=2.1 \times 10^{-4}$. The  surfaces $|A(x,y,z'=0,
t'=0;t'=1)|^2$ are plotted. The transverse size (the spot) grows  by
factor $ \sqrt{2}$ over normalized time-distance $t'=z=1$. This
correspond to real distance $z_0=z_{diff}^{beam}=7.85$ $cm$ equal to
the diffraction length of a laser beam (compare with Fig.1).}
\end{center}
\end{figure}

\begin{figure}[ht]\label{Fig3}
\begin{center}
\includegraphics[width=120mm,height=50mm]{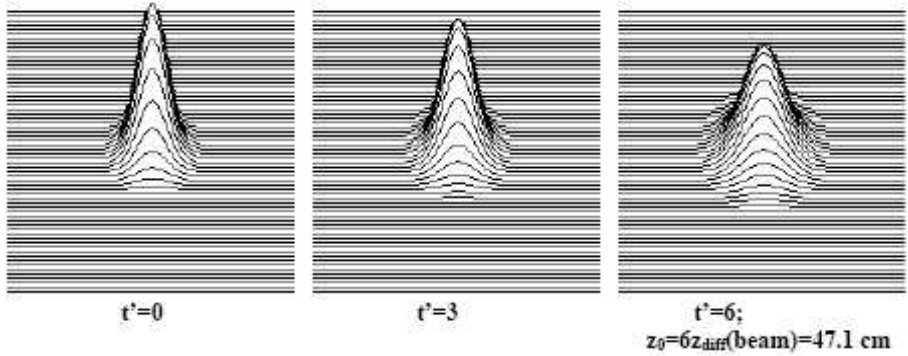}
\caption{Intensity distribution profiles of $43$ $ps$ pulse on
carrying frequency $800$ $nm$ (long Gaussian pulse). Numerical
solutions of the linear SVEA (\ref{lin13}) in Galilean frame is
performed with initial conditions  satisfy $\alpha\delta^2=6$;\
$\alpha=1.\times10^{5}$;\ $\delta^2=6.\times10^{-5}$;\ $\beta_1=2.1
\times 10^{-4}$. The transverse size (the spot) grows  by factor $
\sqrt{2}$ over normalized time-distance $t'=z=6$. This correspond to
a real distance $z_{diff}^{pulse}=\alpha\delta^2
z_{diff}^{beam}=47.1$ $cm$.}
\end{center}
\end{figure}

\subsection{Evolution of long optical pulses (light filaments)}
The pointed above choice for the parameter $\alpha\delta^2=1$ of a
long pulse is used particularly to compare it's diffraction with the
diffraction length of a laser beam. We mark also that in the general
case, the real diffraction length of a long pulse (ns or ps) is
\emph{similar} to the diffraction of laser beam and the difference
is in the factor $\alpha\delta^2$ or:

\begin{eqnarray}
\label{diffpulse}
z^{pulse}_{diff}=\alpha\delta^2z^{beam}_{diff}=k_0^2r^4_{\perp}/z_0.
\end{eqnarray}

The validity of this expression is  illustrated on the next two
figures where the dynamics of an initial long pulse is governed by
the linear SVEA (\ref{lin13}) in Galilean frame with initial
condition $\vec{A}=A_x\vec{x};\
A_x(x,y,z',t'=0)=\exp\left(-\frac{x^2+y^2+z'^2}{2}\right)$ and
dimensionless constants:

a) long pulse on $260$ $ps$: $\alpha\delta^2=1$;\
$\alpha=6.16\times10^{5}$;\ $\delta^2=1/6.16\times10^{-5}$;
$\beta_1=2.1 \times 10^{-4}$.

b) long pulse on $43$ $ps$: $\alpha\delta^2=6$;\
$\alpha=1.\times10^{-5}$;\ $\delta^2=6.\times10^{-5}$; $\beta_1=2.1
\times 10^{-4}$.

Case a) is illustrated on Fig. 2, where the spot (x,y size) of the
pulse is plotted and the parameters  are selected to satisfy the
relation $\alpha\delta^2=1$. That is why  the pulse enlarge its
spatial width by a factor $\sqrt{2}$ on the same normalized
time-distance $t'=z=1$ as in the case of laser beam.  Case b) is
illustrated on Fig. 3., where the important dimmensionless parameter
is $\alpha\delta^2=6$. As can be expected the spot of the pulse
grows by factor $\sqrt{2}$ on six time longer distance than a laser
beam and the special case a). Summarizing the results obtained for
the linear regime of long pulses we conclude: The long optical
pulses admit diffraction length of \emph{the same order} to the one
of a laser beam and this length can be \emph{equal} only in some
partial cases, satisfy $\alpha\delta^2=1$.

\begin{figure}[ht]\label{Fig4}
\begin{center}
\includegraphics[width=120mm,height=50mm]{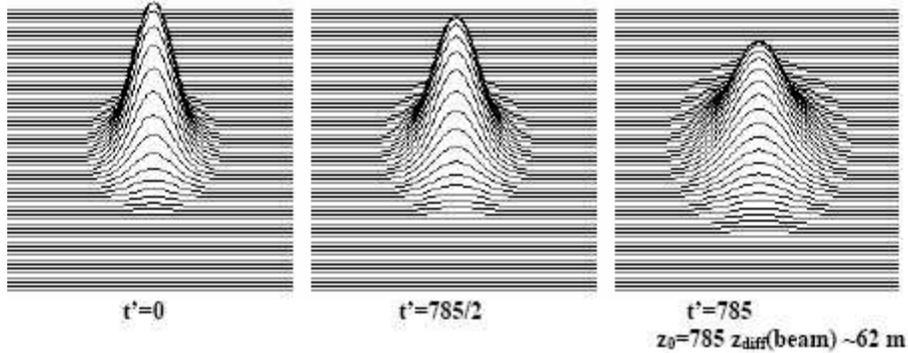}
\caption{Evolution of a Gaussian light bullet with $330$ $fs$ time
duration governed by linear SVEA (\ref{lin13}) in Galilean frame
under initial condition
$A_x(x,y,z,t=0)=\exp\left(-\frac{x^2+y^2+z^2}{2}\right)$,
$\alpha=785.0;\ \delta=1.0$;\ $\beta_1=2.1 \times 10^{-4}$. The
surfaces $|A(x,y,z'=0; t'=0;\ t'=785/2;\ t'=785)|^2$ are plotted.
The transverse size (the spot) grows  by factor $ \sqrt{2}$ over
normalized time-distance $t'=z=785$. For the selected in the paper
laser source this corresponds to a real distance
$z_{diff}^{pulse}=\alpha\delta^2z_{diff}^{beam}\simeq 62$ $m$.}
\end{center}
\end{figure}

\subsection{Propagation of light bullets in linear regime}
The evolution of LB in media with dispersion, is governed by the
same SVEA (\ref{lin13}) as in the case of long pulses. The shape of
the LB is symmetric in the $x$, $y$ and $z$ plane, so that the
linearly polarized initial Gaussian profile can be written as:

\begin{eqnarray}
\label{LBul}
\vec{A}=A_x\vec{x};\ \alpha=785;\  \delta^2=\frac{r_{\bot^2}}{z_0^2}=1,\
\beta_1=2.1 \times 10^{-4}\nonumber\\
A_x(x,y,z,t=0)=\exp\left(-\frac{x^2+y^2+z^2}{2}\right).
\end{eqnarray}

From the qualitative analysis presented in the previous section,
when $\delta^2=1$, the widening of the LB is expected to be
$\alpha=785$; $z^{bullet}_{diff}=\alpha
z^{beam}_{diff}=k_0^2r^{3}_{\perp}$. The surface (x,y plane) of the
solution of VLAE (\ref{L2}) with initial conditions of the kind of
(\ref{LBul}) on  normalized time-distance $t'=z=785$, calculated by
exploit of FFT technique, is illustrated on Fig. 4. One can see that
the pulse enlarge its spot by factor $\sqrt{2}$ along a considerable
distance of $z^{bullet}_{diff}=\alpha z^{beam}_{diff}=785
z^{beam}_{diff}=6162.25$ $cm$ $\cong 61 $ $m$. Let us remark once
again that this result is only correct if the number of harmonics
under the pulse, multiplied by $2\pi$ (dimensionless parameter
$\alpha$), is large.

\begin{figure}[ht]\label{Fig5}
\begin{center}
\includegraphics[width=120mm,height=50mm]{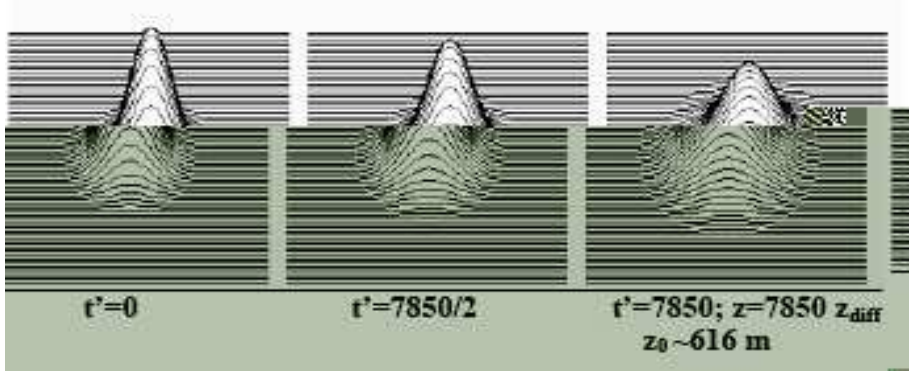}
\caption{Transverse intensity distribution of $33$ $fs$ pulse on
carrying frequency $800$ $nm$ (light disk) governed by the same SVEA
(\ref{lin13}) and initial condition
$A_x(x,y,z,t=0)=\exp\left(-\frac{x^2+y^2+z^2}{2}\right)$,
$\alpha=78.5;\ \delta=100$;\ $\beta_1=2.1 \times 10^{-4}$. The
surfaces $|A(x,y,z'=0; t'=0;\ t'=7850/2;\ t'=7850)|^2$ are
presented. The LD enlarges  its transverse size by factor $\sqrt{2}$
over the normalized time-distance $t'=z=7850$. This correspond to
7850 diffraction lengths of a laser beam or for the  selected laser
source: $z_{diff}^{pulse}=\alpha\delta^2z_{diff}^{beam}\simeq 621$
$m$.}
\end{center}
\end{figure}

\subsection{Dynamics of light disks in linear regime.\\ Low-diffractive regime}
As it was mentioned in the beginning, optical pulses with small
longitudinal and large transverse size, while at the same time the
large number of harmonics under the pulse remaines, can be obtained
without significant experimental difficulties. This can easily be
realized in the optical region for pulses with time duration from
$200-300$ $fs$ up to $30-40$ $fs$. We consider again the propagation
of LD in the framework of the solutions (\ref{LG}) of the SVEA
(\ref{lin13}) in Galilean coordinates under initial conditions of
the form:

\begin{eqnarray}
\label{LLBG}
\vec{A}=A_x\vec{x};\ \alpha=78.5;\  \delta^2=\frac{r_{\bot^2}}{z_0^2}=100, \nonumber\\
A_x(x,y,z',t'=0)=\exp\left(-\frac{x^2+y^2+z'^2}{2}\right).
\end{eqnarray}
Results of calculations of solution (\ref{LG}) with initial
conditions of kind (\ref{LLBG}), using FFT and inverse FFT technique
are presented on Fig. 5. The numerical solution confirm our
expectation that the LD enlarges its shape by factor $\sqrt{2}$ on
7850 time longer distance than the diffraction length of a laser
beam; $z^{disk}_{diff}=\alpha\delta^2
z^{beam}_{diff}=k_0^2r^{4}_{\perp}/z_0=7850z^{beam}_{diff}\cong 616$
$m$. The new formula for diffraction length of optical pulses
(\ref{diffpulse}) gives  remarkable opportunity to select the
parameters of the laser pulse and to obtain pulses with negligible
diffraction. From (\ref{diffpulse}) it is seen that
$z^{pulse}_{diff}$ depends on the spot diameter of the pulse by four
degree ($z^{disk}_{diff}\sim r^{4}_{\perp}$). If we use pulse large
enough in transverse dimension  we can obtain practically
diffractionless pulses. For example, using (\ref{diffpulse}) and
light disk with waist $r_{\perp}=1$ $cm$, time duration $t_0=33$
$fs$ ($z_0=10$ $\mu m$), and $k_0=7.85\times10^4$ $cm^{-1}$
($\lambda_0=800$ $nm$) we can obtain pulse diffraction length of
order of $z^{disk}_{diff}\sim 6160$ $km$. Such LD  will be propagate
in transparency region of gases or vacuum on several thousand
kilometers without practical diffraction enlargement.

\section{Conclusion}
In this paper dynamics of ultrashort laser pulses in media with
dispersion, dispersionless media and vacuum are investigated  in the
frame of non-paraxial generalization of the amplitude equation. In
partial case  of media with dispersion, we obtained an integro -
differential nonlinear equation, governing propagation of optical
pulses with time duration of order of the optical period. The slowly
varying envelope approximation (many harmonics under the pulse)
reduced this amplitude integro - differential equation to the well
known slowly-varying amplitude vector nonlinear differential
equation with different orders of dispersion of the linear and
nonlinear susceptibility. In case of propagation of optical pulses
in dispersionless media and vacuum, we obtained an nonparaxial
amplitude equation which is valid in both cases, namely, pulses with
many harmonics and pulses with only one-two harmonics under the
envelope. We normalized these amplitude equations and obtained five
dimensionless parameters determining different linear and nonlinear
regimes. The nonparaxial envelope equations for media with
dispersion, dispersionless media and vacuum are solved in linear
regime and new fundamental solutions, including the GVD, are found.
In gases and vacuum the solutions of these equations  predict new
diffraction length for optical pulses
$z^{pulse}_{diff}=k_0^2r^4_{\perp}/z_0$. We demonstrate by these
analytical and numerical solutions a significant decreasing of the
diffraction enlargement of $fs$  pulses (LB and LD) in respect to
paraxial widening of a laser beam and a possibility to reach
diffraction-free regime.

\section{Acknowledgements}

This work is partially supported by the Bulgarian Science Foundation
under grant F 1515/2005.

\end{document}